\documentstyle[multicol,prl,aps,epsfig]{revtex}

\newcommand{\be}{\begin{equation}}
\newcommand{\ee}{\end{equation}}
\newcommand{\ba}{\begin{eqnarray}}
\newcommand{\ea}{\end{eqnarray}}
\newcommand{\baa}{\begin{eqnarray*}}
\newcommand{\eaa}{\end{eqnarray*}}
\newcommand{\bb}{\begin{thebibliography}}
\newcommand{\eb}{\end{thebibliography}}

\newcommand{\bm}[1]{\mbox{\boldmath $#1$}}
\newcommand{\ci}[1]{\cite{#1}}
\newcommand{\bi}[1]{\bibitem{#1}}

\input colordvi

\begin {document}

\title{Spin-independent and double-spin {\mbox{\boldmath{$\cos\phi$}}}
asymmetries in semi-inclusive pion electroproduction\\ 
}

\author{
K.A.~Oganessyan$^{a,b}$, 
L.S.~Asilyan$^a$,  
M.~Anselmino$^c$,
E.~De~Sanctis$^a$
}

\address{
$^a$INFN-Laboratori Nazionali di Frascati I-00044 Frascati, 
via Enrico Fermi 40, Italy \\
$^b$DESY, Deutsches Elektronen Synchrotron 
Notkestrasse 85, 22603 Hamburg, Germany \\
$^c$Dipartimento di Fisica Teorica, Universit\`a di Torino and \\ 
INFN, Sezione di Torino, Via P. Giuria 1, I-10125 Torino, Italy
}

 \maketitle
%
\begin{abstract}
We consider the $\cos\phi$ dependence of the longitudinal double-spin 
asymmetry for charged pion electroproduction in semi-inclusive deep 
inelastic scattering, emphasizing intrinsic transverse momentum 
effects. This azimuthal asymmetry allows to measure the $\cos\phi$ moments
of the unpolarized and double-spin cross-section, simultaneously. 
The size of the asymmetry, in the approximation where all twist-3 
interaction-dependent distribution and fragmentation functions are set 
to zero, is estimated for HERMES kinematics; both the spin-independent and 
the double-spin $\cos\phi$ moments are predicted to be sizable and negative.
\\\newline
PACS numbers: 13.87.Fh, 13.60.-r, 13.88.+e, 14.20.Dh
\end{abstract}
%
%
%
%

\begin{multicols}{2}[]
%

In the context of asymptotically free quantum chromodynamics (QCD) quarks 
and gluons should be produced as free particles. However, as they are 
not observed, non-perturbative confining effects become crucial in the 
amount of partons in the initial state and the formation of hadrons in 
the final state. Such effects, at present, cannot be calculated from first 
principles. They are parameterized in an effective way by introducing 
longitudinal and transverse (the so-called ``intrinsic'' transverse 
motion) degrees of freedom in the parton distribution and fragmentation 
functions. One of the most interesting consequences of non-zero intrinsic 
transverse momentum of partons in hadrons is the non trivial azimuthal 
dependence of the cross-sections for hadron production in hard scattering 
processes.
  
We focus on semi-inclusive deep inelastic scattering (DIS), $eN \to ehX$.
Defining a coordinate system in the laboratory frame with the $z$ axis along 
the momentum transfer $\bm q = \bm k_1 - \bm k_2$ between the initial and 
final lepton and the $x$ axis in the leptonic plane, the 
component of the detected hadron momentum transverse to $\bm q$, 
$\bm P_{h\perp}$, and its azimuthal orientation, $\phi$ (see Fig.1), 
provide interesting variables to study non-perturbative~\ci{CAHN,TM,AK} and 
perturbative effects~\ci{PG,HAG}. Recently, a particular $\cos\phi$ 
moment of the polarized cross-section has been considered~\ci{OMD} and a 
sizable asymmetry, in a Wandzura-Wilczek (WW) like approximation for 
$\pi^{+}$ electroproduction, was predicted. 

We consider here the $\cos\phi$ azimuthal dependence of the usual longitudinal 
double-spin asymmetry: 
\begin{equation}
A_{LL} = \frac{d\sigma^{++}+d\sigma^{--}-d\sigma^{+-}-d\sigma^{-+}}
{d\sigma^{++}+d\sigma^{--}+d\sigma^{+-}+d\sigma^{-+} } \>,
\label{AS}
\end{equation}
where the subscript $LL$ denotes the longitudinal polarization of the beam 
and target respectively and $d \sigma$ is a shorthand notation for  
${d \sigma^{eN \to e h X} /{dx\,dy\,dz\,d^2\bm P_{h\,\perp}}}$; 
the superscripts $++,-- \, (+-,-+)$ denote the helicity states of the beam 
and target respectively, corresponding to antiparallel (parallel) 
polarization\footnote{It leads to a positive $g_1(x)$. }, and   
$x$, $y$, and $z$, are the standard leptoproduction variables defined as   
$$
x=\frac{Q^2}{2(P \cdot q)}, \quad y=\frac{P \cdot q}{P \cdot k_1}, 
\quad z=\frac{P \cdot P_h}{P \cdot q}, 
$$ 
where $k_1$ and $P_h$ are the four-momenta of the incoming charged 
lepton and of the observed hadron, respectively.  

\begin{figure}[htb]
\epsfxsize 6.0 cm {\epsfbox{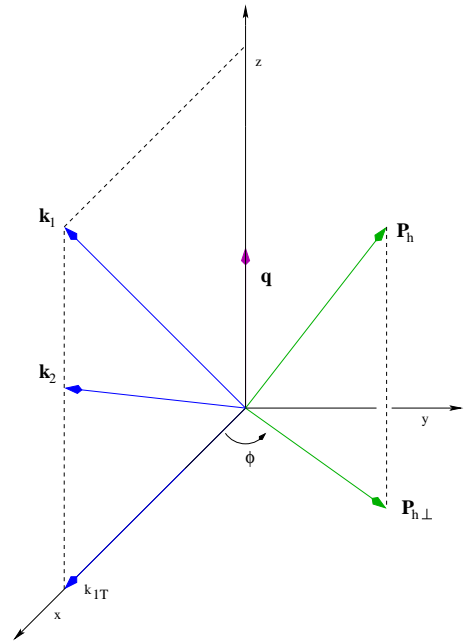}}
\caption{The kinematics of semi-inclusive DIS. }
\label{f1}
\end{figure} 

As it will be shown, the most interesting consequence of this asymmetry 
is that it allows to determine both the spin-independent and double-spin 
$\cos\phi$ moments of the cross-section, simultaneously, without any 
dilution from geometrical acceptance of the spectrometer. 

Due to parity conservation of the electromagnetic and strong interactions, 
Eq. (\ref{AS}) can be written in terms of the spin-independent ($\sigma_{UU}
\equiv (d\sigma^{++}+d\sigma^{--}+d\sigma^{+-}+d\sigma^{-+})/4$) 
and double-spin ($\Delta \sigma_{LL} \equiv (d\sigma^{++} + d\sigma^{--} -
d\sigma^{+-} - d\sigma^{-+})/2$) cross-sections for semi-inclusive DIS, 
\be
\label{AS1}
A_{LL} =\frac{\Delta \sigma_{LL}}{2 \, \sigma_{UU}} \>, 
\ee
where, in general, 
\be
\label{AS2}
\Delta \sigma_{LL} = \sum_{m=1} \Delta \sigma^{m}_{LL} \cos {([m-1] 
\cdot \phi)} \>, 
\ee
\be
\label{AS3}  
\sigma_{UU} = \sum_{m=1} \sigma^{m}_{UU} \cos {([m-1] \cdot \phi)}\>.  
\ee
The subscripts $U$ and $L$ stand for unpolarized and longitudinally 
polarized beam and target. At sub-leading $(1/Q)$ order one has non-vanishing 
spin-independent and double-spin $\cos\phi$ asymmetries: they originate from
non-perturbative effects, both kinematical~\cite{CAHN,AK,OMD} and
dynamical~\ci{TM}, and from perturbative~\ci{PG} effects.
A $\cos2\phi$ asymmetry only appears at order $1/Q^2$~\cite{CAHN,OABD},
unless one allows for time-reversal odd structure functions \cite{BM}. 
We do not consider such contributions here. 
Then, up to sub-leading order $1/Q$, Eq.(\ref{AS1}) can be rewritten as  
\be
A_{LL} = \frac{\Delta \sigma^{1}_{LL}/2 \sigma^{1}_{UU} + 
{\langle \cos\phi \rangle}_{LL} 
\cdot \cos\phi}{1 + 2\,{\langle \cos\phi \rangle}_{UU} \cdot \cos\phi } ,  
\label{AP}
\ee
where ${\langle\cos\phi\rangle}_{UU}$ and ${\langle\cos\phi\rangle}_{LL}$ 
are the unpolarized and double polarized $\cos\phi$ moments, respectively:  
\ba
\label{MOM1}
{\langle \cos\phi \rangle}_{UU} \equiv 
\frac{\int d\phi \cdot \sigma_{UU} \cdot \cos\phi} 
{\int d\phi \cdot \sigma} = \frac{\sigma^{2}_{UU}}{2\,\sigma^{1}_{UU}}\>, \\
\label{MOM2} 
{\langle \cos\phi \rangle}_{LL} \equiv 
\frac{\int d\phi \cdot \Delta \sigma_{LL} \cdot \cos\phi }
{\int d\phi \cdot \sigma} = \frac{\Delta \sigma^{2}_{LL}}
{2\, \sigma^{1}_{UU}} \> \cdot 
\ea
Eq. (\ref{AP}) shows how a measurement of the $\cos\phi$ dependence of 
$A_{LL}$ allows a determination of the moments (\ref{MOM1}) and (\ref{MOM2}) 
($\Delta \sigma^{1}_{LL}$ and $\sigma^{1}_{UU}$ are given by the usual
collinear partonic expressions). In order to see whether or not such a 
$\cos\phi$ dependence is significant and detectable, we give an estimate 
of $A_{LL}(\cos\phi)$ at HERMES energies. To this purpose we need 
to evaluate the $\cos\phi$-moments and we do it by proceeding in the same 
way as in Ref.~\cite{OMD}, {\it i.e.} we use the approximation that all 
interaction-dependent functions are equal to zero. The explicit formulas 
corresponding to this approximation are given in Ref.~\cite{OMD}. 

Let us consider the detector acceptance effects in the $\cos\phi$ weighted 
$A^{\cos\phi}_{{(LL)}_{lab}}$ asymmetry defined as~\cite{OMD} 
\begin{equation}
A^{\cos\phi}_{{(LL)}_{lab}} = \frac{\int d\phi \cos\phi \cdot  
\Delta\sigma_{LL}(\phi)}{{1 \over 4} \int d\phi \cdot \sigma_{UU}(\phi)}. 
\label{E0}
\end{equation}   
Here, the measured polarized (unpolarized) cross section for 
semi-inclusive DIS is the product 
of $\Delta\sigma_{{LL}} \cdot \epsilon(\phi)$ 
($\sigma_{{UU}} \cdot \epsilon(\phi)$), 
where $\epsilon(\phi)$ is the detector acceptance. It can be 
expanded in Fourier series,  
\begin{equation}
\epsilon(\phi) = C_0+\sum_{m=1} \bigg [ C_m \cos(m\cdot \phi)
+D_m \sin(m \cdot \phi) \bigg ]. 
\label{E1}
\end{equation}    
Using the Eqs.(\ref{AS2}) and (\ref{AS3}) and integrating 
over $\phi$ one obtains (up to sub-leading $1/Q$ order) 
\begin{equation}
A^{\cos\phi}_{{(LL)}_{lab}}=4\frac{C_1\cdot\Delta\sigma^1_{LL}
+C_0\cdot\Delta\sigma^2_{LL}+{1 \over 2} C_2\cdot\Delta\sigma^2_{LL}}
{2C_0\cdot \sigma^1_{UU}+C_1\cdot\sigma^2_{UU}}. 
\label{MM}
\end{equation} 
So, the asymmetry clearly depends on the detector acceptance.  
On the contrary, the acceptance effects, being independent of beam and 
target helicities, cancel out in the ratio which gives 
the cross section azimuthal asymmetry of Eq.(\ref{AS}). It is worth 
noticing that the $\phi$-integrated asymmetry $A_{LL}$ is also sensitive 
to the acceptance, which may affect the determination and interpretation  
of $A_1$ in semi-inclusive DIS (let us denote it by $A^h_1$): 
\begin{equation}
A^{h}_1 \equiv A_{LL} = \frac{\int d\phi \cdot \Delta\sigma_{LL} 
\cdot \epsilon(\phi)}{\int d\phi \cdot \sigma_{UU} 
\cdot \epsilon(\phi)}, 
\label{SIH1}
\end{equation}   
which after the integration over $\phi$ becomes (again up to 
sub-leading $1/Q$ order), 
\begin{equation}
A^{h}_1 = \frac{\Delta\sigma^1_{LL}/\sigma^1_{UU}
+\frac{C_1}{C_0}\cdot{\langle \cos\phi \rangle}_{LL}}
{1 + \frac{C_1}{C_0}\cdot{\langle \cos\phi \rangle}_{UU}}. 
\label{SIH2}
\end{equation}  
Therefore, aside from their own physical interest, the $\cos\phi$  
moments must be carefully determined in order to take into account
their effects on the $\phi$-integrated semi-inclusive asymmetries $A^h_1$.    
    
In Figs.~2 and 3, the unpolarized and doubly polarized $\cos\phi$ moments 
defined by Eqs. (\ref{MOM1}) and (\ref{MOM2}) for $\pi^{+}$ production on 
a proton target are shown as functions of $x$ for three different values 
of the mean squared transverse momentum $\langle p_T^2 \rangle$ of the 
initial parton.     
\begin{figure}[htb]
\epsfxsize 7.0 cm {\epsfbox{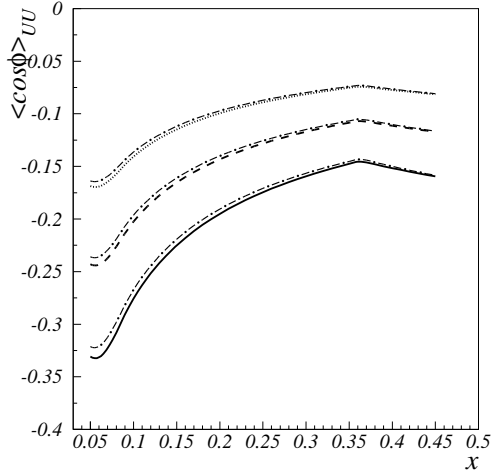}}
\caption{
The ${\langle \cos\phi \rangle}_{UU}$ moment as a function of $x$, at 
$\langle p^2_T \rangle = (0.5)^2$~GeV$^2$ (dotted-curve),   
$\langle p^2_T \rangle = (0.6)^2$~GeV$^2$ (dashed-curve), and 
$\langle p^2_T \rangle = (0.7)^2$~GeV$^2$ (full-curve), 
calculated using the parameterizations of Refs.~\protect \cite{BBS,REYA} 
(full-curves). The dash-dotted curves are obtained using the 
sets of LO parameterizations of Refs.~\protect \cite{GRV,AAC,KKP} }. 
\label{f2}
\end{figure}
\begin{figure}[htb]
\epsfxsize 7.0 cm {\epsfbox{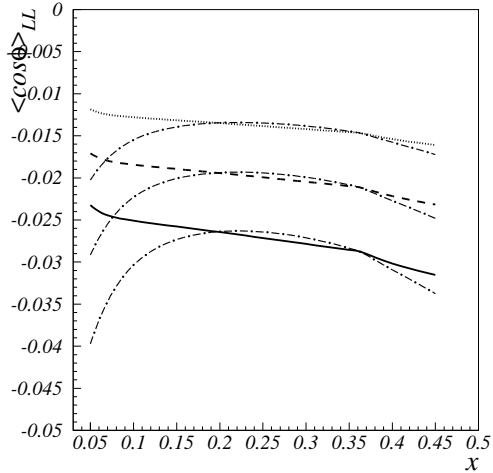}}
\caption{
The ${\langle \cos\phi \rangle}_{LL}$ moment as a function of $x$. 
The notations for the curves are the same as in Fig.2. }
\label{f3}
\end{figure}
The curves are calculated by integrating over HERMES kinematical ranges, 
corresponding to $1$ GeV$^2$ $\leq Q^2 \leq 15$ GeV$^2$, 
$4.5$ GeV $\leq E_{\pi} \leq 13.5$ GeV, $0.2 \leq z \leq 0.7$, 
and $0.2 \leq y \leq 0.8$ and taking $\langle P_{h\perp} \rangle = 0.4$ 
GeV as input~\cite{JGO}. We use $Q^2$-independent 
parameterizations for the distribution, 
$f_1(x), g_1(x)$~\cite{BBS}, and fragmentation, $D_1(z)$~\cite{REYA}, 
functions, as well as the recent sets of LO distribution~\cite{GRV,AAC} 
and fragmentation~\cite{KKP} functions. From Fig.~2 one can see that 
the result for spin-independent azimuthal asymmetry is insensitive 
to the input parameterizations choice due to the cancellation of effects 
in the ratio defining ${\langle \cos\phi \rangle}_{UU}$, while  
double-spin asymmetry (dash-dotted curves in Fig.~3) displays essential 
changes at $x < 0.2$. Note that the change in behavior of the 
curves at $x$ around 0.36 is only due to the integrations of chosen 
kinematical ranges. 

The figures show that, within WW approximation, both 
spin-independent and double-spin $\cos\phi$ moments are negative 
and large enough in magnitude to be measurable (in particular the 
spin-independent one) at HERMES. 
The results for $\pi^-$ are similar and only slightly smaller. 
The numerical results for ${\langle \cos\phi \rangle}_{LL}$ given in 
Fig.~3 and of $A^{\cos\phi}_{{(LL)}_{lab}}$ given in Fig.~2 
of Ref.~\cite{OMD} are consistent if one takes into account that 
${\langle \cos\phi \rangle}_{LL} \simeq 
{ \langle P_{h\perp} \rangle \over 8} \cdot A^{\cos\phi}_{{(LL)}_{lab}}$, 
the choice of the different $\langle P_{h\perp} \rangle$, and the 
linear dependence on $\langle p^2_T \rangle$. The ``kinematical'' 
contribution to ${\langle \cos\phi \rangle}_{LL}$ 
coming from the transverse component of the target polarization is  
small~\cite{OMD} and was not taken into account.

Our estimates also indicate that $\cos\phi$-moments are very sensitive to 
the $\langle p_T^2 \rangle$ value. This underlines the importance of 
having a reliable value of $\langle p_T^2 \rangle$, at least for  
the considered kinematical regions. On the other hand, the measurement 
of the moments, via Eq. (\ref{AP}), may provide an estimate of 
$\langle p^2_T \rangle$. 
\begin{figure}[htb]
\epsfxsize 7.0 cm {\epsfbox{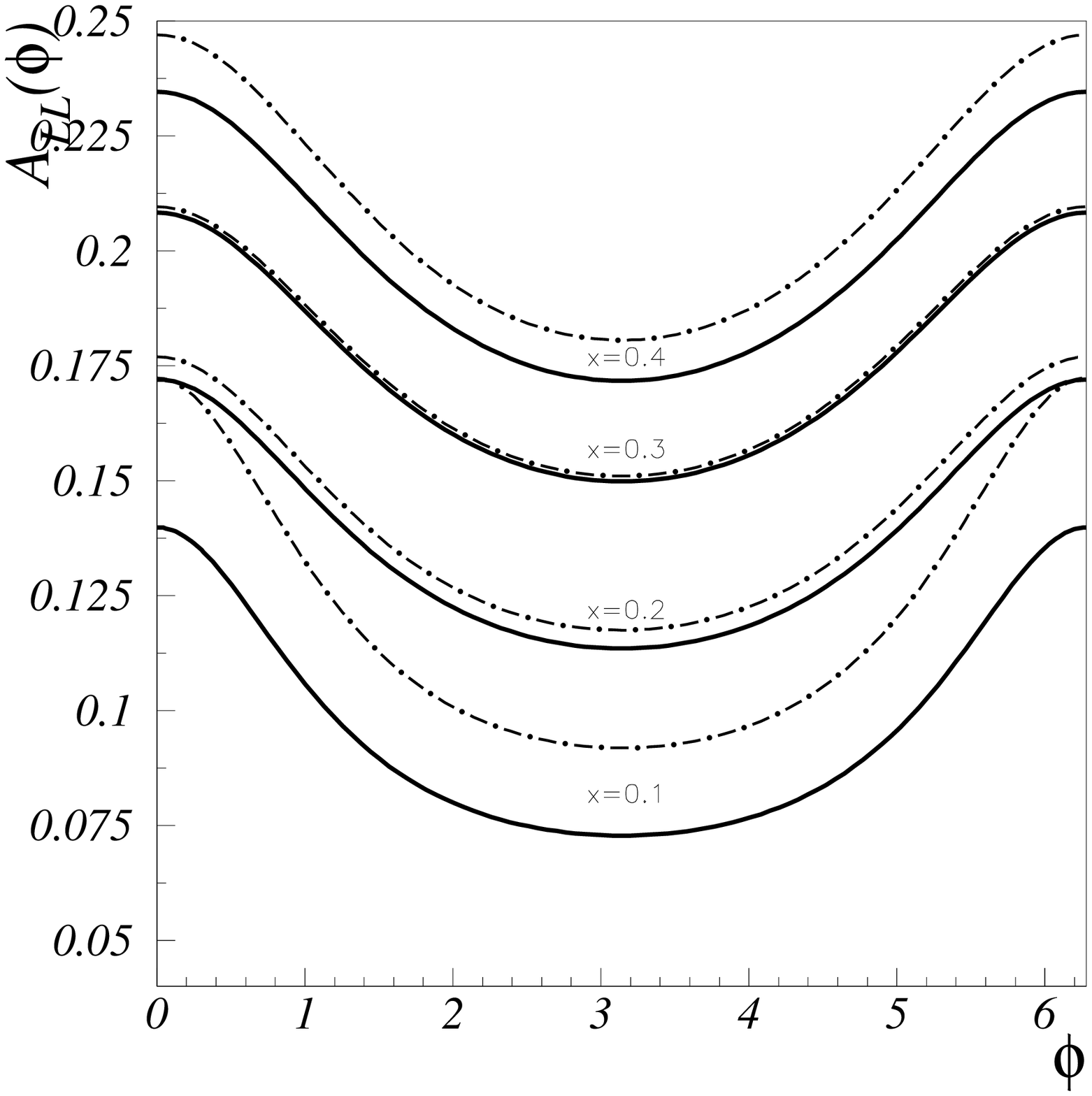}}
\caption{The $A_{LL}$ asymmetry for $\pi^{+}$ 
production as a function of azimuthal angle $\phi$ for different values of 
Bjorken $x$ at $\langle p^2_T \rangle = (0.7)^2$~GeV$^2$, 
The notations for the dash-dotted curves are the same as in Fig.~2.} 
\label{f4}
\end{figure}   
The average value of the $P^2_{h\perp}$ distribution, assuming that $p_T$ and 
$k^{'}_T$ are independent of kinematical variables, is given by~\cite{MRS}
\be
\label{PT2}
\langle P^2_{h\perp} \rangle = {\langle z^2 \rangle}{\langle p_T^2 
\rangle}_{(Q^2,x)}  + {\langle k^{'2}_T \rangle}_{(Q^2,z)},
\ee
where $k^{'}_T$ is the transverse momentum of the produced hadron with 
respect to the quark momentum. The subscripts on the right-hand side 
recall the variables which the averages generally depend on. This  
does not allow to consider the $\cos\phi$ asymmetry 
consistently at different kinematical conditions and obtain some constraints 
on mean transverse momenta from available data~\cite{EMC,E665}. In the 
considered kinematical regions, $\langle z^2 \rangle$ is small 
($\sim 0.2$) and then its contribution is strongly suppressed. Then 
$k^{'}_T$ gives the main contribution to $\langle P^2_{h\perp} \rangle$. 
Using the values $\langle k^{'2}_T \rangle = (0.44)^2$ GeV$^2$~\cite{PYTHIA} 
and $\langle p_T^2 \rangle = (0.5)^2, (0.6)^2, (0.7)^2$ GeV$^2$, the 
corresponding values of $\langle P^2_{h\perp} \rangle $ are $0.24$, $0.265$, 
and $0.29$ GeV$^2$, respectively; they are reasonable for HERMES kinematics. 
It is worth noticing that these values of the parameters lead to  
the same results for $\cos\phi$ moments if one integrates over 
$P_{h\perp}$ assuming a Gaussian transverse momentum dependence of 
distribution and fragmentation functions. 

In Fig.~4 the $A_{LL}$ asymmetry of Eq.(\ref{AP}) for $\pi^{+}$ production 
on a proton target is presented as a function of the azimuthal angle 
$\phi$ at $\langle p^2_T \rangle = (0.7)^2$~GeV$^2$, and for different 
values of $x$. The strong dependence of the magnitudes of the 
double-spin $\cos\phi$ and $\phi$-independent asymmetries on the 
input parameterizations at low x-region leads to the changes of 
$A_{LL}$ asymmetry shown in Fig.~4. As it is seen, the asymmetry 
is large and well detectable experimentally. Hence, it can 
provide the simultaneous measurement of the unpolarized and double 
polarized $\cos\phi$ moments.    

In conclusion, we have considered effects related to parton intrinsic motion,
which, in semi-inclusive DIS processes, manifest as azimuthal dependences 
of cross-sections; in particular, we have examined the $\cos\phi$ dependence
of the double longitudinal spin asymmetry for charged pion electroproduction.
This dependence has been shown to be measurable; it depends on and allows the 
simultaneous determination of the spin-independent and double-spin 
$\cos\phi$-moments of the cross-section. 
    
The sizes of these moments, in the approximation where all twist-3 
interaction-dependent distribution and fragmentation functions are set to 
zero, is estimated for HERMES kinematical configurations; they turn out 
to be significantly large in magnitude and negative. The results also 
indicate a great sensitivity to the choice of the partons intrinsic 
average transverse momentum; they might provide a way of access and an 
estimate for the value of $\langle p^2_T \rangle$.  

The proposed asymmetry is measurable in running or planned experiments 
at HERMES, COMPASS, JLAB (upgraded) and may answer the question of the 
importance of twist-3 contributions in semi-inclusive DIS as well as 
provide some information on the $p_T$ behavior of the structure functions 
$f_1(x)$ and $g_1(x)$.   

We thank A.~Bacchetta, D.~Boer, U.~D'Alesio, R.L.~Jaffe, R.~Jakob, 
A.~Kotzinian, R.~Kundu, A.~Metz,  P.J.~Mulders, D.~M\"uller, and 
A.~Sch\"afer for many useful discussions. Discussions with N.~Bianchi, 
N.~Makins, A.~Miller, and V.~Muccifora are also acknowledged.   

\bb{99}

  \bi{CAHN} R.N. Cahn, Phys. Lett.  B {\bf 78} (1978) 269; Phys. Rev. D 
        {\bf 40} (1989)
  \bi{TM} P.J.~Mulders and R.D.~Tangerman, Nucl Phys. B {\bf 461} (1996) 197.
  \bi{AK} A.~Kotzinian, Nucl. Phys. B {\bf 441} (1995) 234.
  \bi{PG} H.~Georgi, H.D.~Politzer, Phys. Rev. Lett. {\bf 40} (1978) 3.
  \bi{HAG} K.~Hagiwara, K.~Hikasa, N.~Kai, Phys. Rev. D {\bf 27} (1983) 84. 
  \bi{OMD} K.A.~Oganessyan, P.J.~Mulders, E.~De~Sanctis, Phys. Lett. 
           {\bf B532} (2002) 87.  
  \bi{OABD} K.A.~Oganessyan, et.al., Eur. Phys. J. {\bf C 5} 681 (1998).
  \bi{BM} D.~Boer, P.J.~Mulders, Phys.Rev. D {\bf 57} (1998) 5780.
  \bi{JGO} A.A.~Jgoun, hep-ex/0107003.
  \bi{BBS} S.~Brodsky, M.~Burkardt, I.~Schmidt, Nucl. Phys. {\bf B441} 
           (1995) 197. 
  \bi{REYA} E.~Reya, Phys. Rep. {\bf 69} (1981) 195.
  \bi{GRV} M.~Gluck, E.~Reya, A.~Vogt, Z. Phys. C {\bf 67} (1995) 433. 
  \bi{AAC} Y.~Goto, et. al, Phys. Rev. D {\bf 62} (2000) 034017. 
  \bi{KKP} B.A.~Kniehl, G.~Kramer, B.~P\"{o}tter, Nucl. Phys. B {\bf 582} 
        (2000) 514.       
  \bi{MRS} A.~Mendez, A.~Raychaudhuri, V.J.~Stenger, Nucl. Phys. 
           {\bf B148} 499, (1979). 
   \bi{EMC} M. Arneodo et al. (EMC Collaboration),  Z. Phys. 
        C {\bf 34} (1987) 277.  
   \bi{E665} M.R. Adams et al. (E665 Collaboration), 
        Phys. Rev. D {\bf 48} (1993) 5057.  
  \bi{PYTHIA} T.~Sjostrand, et.al, Comput. Phys. Commun. 
              {\bf 135} (2001) 238; [hep-ph/0108264].  
  
\eb

\end{multicols}

\end{document}